\documentclass[pre,twocolumn]{revtex4}

\usepackage{amsmath,amssymb}    
\usepackage{amsfonts}            

\usepackage{graphicx}           

\newcommand*{\eg}{e.\,g.}
\newcommand*{\ie}{i.\,e.} 
\newcommand*{\dor}{\op{\varrho}}

\providecommand*{\op}{\hat}

\newcommand*{\kom}[2]{[#1,#2]}
\newcommand*{\Kom}[2]{\left[ #1,#2\right]}

\newcommand*{\komm}[2]{\bigl[ #1,#2\bigr]}

\newcommand*{\Aop}{\op{A}}
\newcommand*{\Bop}{\op{B}}

\newcommand*{\ord}[1]{O\hspace{-1pt}(#1)}

\newcommand*{\ee}{\mathrm{e}}
\newcommand*{\dd}{\mathrm{d}}
\newcommand*{\iu}{\mathrm{i}}
\newcommand*{\Hop}{\op{H}}
\newcommand*{\dkommAneu}{\Kom{\vphantom{P^P}A}{\kom{A}{B}}}
\newcommand*{\dkommBneu}{\Kom{\vphantom{P^P}B}{\kom{A}{B}}}

\begin{document}
\title{Model Studies on the Quantum Jarzynski Relation}
\author{Jens Teifel and G\"{u}nter Mahler}
\affiliation{Institut f\"{u}r Theoretische Physik 1, Universit\"{a}t
  Stuttgart, 70550 Stuttgart, Germany}
\date{\today}
\begin{abstract}
  We study the quantum Jarzynski relation for driven quantum models
  embedded in various environments. We do so by generalizing a proof
  presented by Mukamel [Phys. Rev. Lett. \textbf{90}, 170604 (2003)] for
  closed quantum systems. In this way, we are able to prove that the
  Jarzynski relation also holds for a bipartite system with
  microcanonical coupling. Furthermore, we show that, under the
  assumption that the interaction energy remains constant during the
  whole process, the relation is valid even for canonical coupling.
  The same follows for open quantum systems at high initial
  temperatures up to third order of the inverse temperature. Our
  analytical study is complemented by a numerical investigation of a
  special model system.
\end{abstract}
\maketitle
\section{Introduction}
If a system is driven far away from thermal equilibrium by an external
force, its behavior can no longer be described by linear response
theory or other near-equilibrium approximations. Fluctuations may
dominate the evolution of the system. Recently, some astonishingly
general theorems, which make exact statements, have been
found~\cite{prl_jarzynski_nonequilibriumequality,jsp_Jarzynski_detailedft,prl_evans_probabilitysecondlawviolation,pre_crooks_entropyproductionft,prl_seifert_entropyproduction}.
For classically described systems there exist several proofs for
various kinds of Hamilton
dynamics~\cite{jsp_Jarzynski_detailedft,prl_seifert_entropyproduction}.

One faces a different situation for systems the dynamics of which are
described by quantum mechanics. Mostly for very restricted situations
or only for very special model systems have those theorems been shown
to
hold~\cite{arxiv_kurchan_quantumft,pre_monnai_unifiedqftandjr,tecnote_tasaki_qmsystemsapplications,arxiv_monnai_quantumcorrectionft,pre_allahverdyan_quantumworkfluctuation,prl_classical_qm_FT_heatexchange,{pre_quantum_version_free_energy-work}}
so far. However, there exists a general proof of the Jarzynski
relation, a non-equilibrium work theorem, for closed quantum systems
\cite{prl_mukamel_quantumjarzynski}. Generalizations to open quantum
systems have only been discussed for restricted model systems or various
constraints on the microscopic
dynamics~\cite{pre_quantum_masterequation_jarzynski,prl_classical_qm_FT_heatexchange}.

One major problem is that the proofs of the so-called
group of non-equilibrium fluctuation theorems as well as the work
fluctuation theorems relate properties of trajectories from forward
and backward processes with each other. Trying to transfer these
proofs to quantum systems brings up the problem of trajectories in the
quantum case.

Another, possibly even more severe problem is the question of what
exactly has to be understood by fluctuations in a quantum system. In
quantum mechanics the state of a system is fully described by its
density operator. Without any measurement it is ambiguous to speak
about fluctuations, because the probability of a measurement outcome
of a certain quantity might vary, but not the untouched quantity
itself.  Fluctuations of one-time quantities are not ambiguous as far
as their measurement is concerned. If, however, the relevant
observable of two-time quantities, like work, \eg, does not commute
with itself at different times then even the measurement of
fluctuation becomes ambiguous.  Every measurement changes the state of
the system thereby influencing its evolution. In the case of work
fluctuation theorems the measurement has naturally to be included
since one needs to measure the amount of work performed on the system.
This measurement will then yield different values for each repetition
and is thus said to fluctuate.  So, in this sense we have no problem
speaking of fluctuations.

This is why we try the following ansatz: We concentrate on a work
fluctuation theorem. In order to circumvent the problem of the
definition of a quantum trajectory, we use the quite general proof for
closed quantum systems~\cite{prl_mukamel_quantumjarzynski}, and try to
generalize it to open quantum systems without having to define
any quantum trajectory (in the sense of stochastic unravelling, \ie\
continuous measurement). This will be done by splitting up the
exponential function of the partition sum into its system and
environmental part, respectively. There is no need to specify the
underlying dynamics any further. It is not necessary to measure the
system during the process since we rely on measurements at the
beginning and the end of the process only. The idea of a two-time
measurement approach has first been developed by
Kurchan~\cite{arxiv_kurchan_quantumft}.

First, we start by a brief review of the Jarzynski relation in
section~\ref{jarzynski_short}. Then, in section~\ref{analytic} we will
discuss conditions under which we will prove that the Jarzynski relation is
valid. This analytical study 
will then be supplemented by a discussion of our numerical
results in section~\ref{numeric}.

\section{Jarzynski relation \label{jarzynski_short}}
If we perform a process on a system initially in a canonical state,
the system will, in general, be driven out of equilibrium. Therefore,
the average work $\overline{W}$ we have to bring up in order to
perform the process will exceed the free energy difference $\Delta F$
between initial and final state:
\begin{equation}
\overline{W} \geq \Delta F.  
\end{equation}
In 1997, C. Jarzynski came up with a remarkable relation that connects
the non-equilibrium variable $W$ with an equilibrium property of the
system, $\Delta F$~\cite{prl_jarzynski_nonequilibriumequality}:
\begin{equation}
\overline{\ee^{-\beta W}} = \ee^{-\beta\Delta F} = \frac{Z(t)}{Z(0)}.
\label{jarzynski} 
\end{equation}
This relation holds no matter how far the system is driven out of
equilibrium. It allows to access the free energy difference between
two states via non-equilibrium measurements. This is very useful for
experiments which cannot be carried out in
quasi-equilibrium~\cite{arxiv_ritort_work_fluctuations}: Note that the
right-hand side does not depend on how we get from the initial to the
final state.

The Jarzynski relation has been generalized to closed quantum systems
by S. Mukamel~\cite{prl_mukamel_quantumjarzynski} under the condition,
that the system is initially in a canonical state.

\section{Analytical Results \label{analytic}}
We consider a bipartite system, split into the so-called system and
environment, respectively. An arbitrary process is performed on the
system leading to an explicitly time-dependent Hamilton operator. The
compound system is considered closed and the complete Hamilton operator reads 
\begin{equation}
\Hop (t) = \Hop^S(t)\otimes\op{1}^C + \op{1}^S\otimes\Hop^C + \sigma
\Hop^\text{int}, \label{def_hamilton}
\end{equation}
where $\Hop^S(t)$ denotes the Hamiltonian of the system, $\Hop^C$ the
operator for the environment and 
$\Hop^\text{int}$ the interaction Hamiltonian with the coupling
constant $\sigma$, determining the strength of the coupling. In the
following, we investigate different coupling scenarios.  

\subsection{Microcanonical Coupling}

\begin{figure}
  \begin{center}  \scalebox{0.65}{
    \includegraphics[scale=0.9]{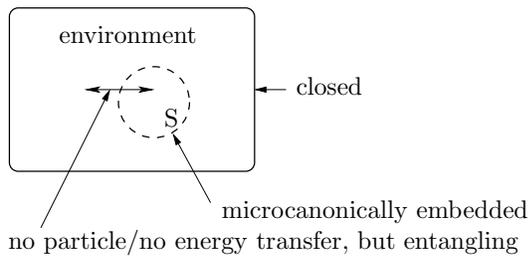}
    
    }
    \caption{Microcanonical coupling: between a small subsystem S and
      its environment. Total system is a closed compound system,
      initially in the canonical state
      $\op{\varrho}(0)=\frac{\ee^{-\beta\Hop(0)}}{Z(0)}$
      \label{sys_env_model_micro}} 
  \end{center}
\end{figure}

We consider a microcanonical
coupling~\cite{book_gemmer_quantum_thermodynamics} (cf.
Fig.~\ref{sys_env_model_micro}) first. This means that neither
particle nor energy exchange between the system and the environment is
allowed.  Note that because of entanglement between the system and the
environment the work performed on the system will, nevertheless, be
influenced by the environment. This is due to the fact that the
off-diagonal elements of the reduced density operator of the system
will be damped away. Since these elements are crucial for transitions
within the system, the work should, indeed, be influenced by the
microcanonical coupling.

Since no energy transfer between system and environment is allowed, we
require that at any time $t$ the Hamilton operator of the
environment and the system, respectively, have to be separate
constants of motion: 
\begin{equation}
\kom{\Hop(t)}{\Hop^C} = 0 \text{, }\; \kom{\Hop(t)}{\Hop^S(t)} = 0.
\end{equation}
Using $\kom{\Hop^S(t)}{\Hop^C}$ = 0 we readily arrive at:
\begin{equation}
\kom{\Hop^C}{\Hop^\text{int}}=0  \text{,
}\;\kom{\Hop^S(t)}{\Hop^\text{int}}= 0. \label{kom_microcanonical}
\end{equation}
The quantum Jarzynski relation for closed systems states that
[cf.~(\ref{jarzynski})]
\begin{equation}
\overline{\ee^{-\beta W}} = \frac{Z(t)}{Z(0)} \label{eq:jarzynski_2}
\end{equation}
holds for arbitrary processes. Our aim is to prove that this relation
also holds for this type of open quantum systems, \ie, for
microcanonical coupling. To this end, we investigate the relation of
the partition sums:
\begin{equation}
Z(t)=\text{Tr}\!\left[\ee^{-\beta\Hop(t)} \right] =
\text{Tr}\!\left[\ee^{-\beta(\Hop^S(t)\otimes\op{1}^C +
    \op{1}^S\otimes \Hop^C + \Hop^I)} \right].
\end{equation}
Here we have introduced the abbreviation $\Hop^I\equiv
\sigma\Hop^\text{int}$.  Since the Hamilton operators commute with
each other according to~(\ref{kom_microcanonical}), we can split the
exponential function, arriving at
\begin{equation}
Z(t)= \text{Tr}\!\left[\ee^{-\beta\Hop^S(t)}\otimes \ee^{-\beta\Hop^C}
  \ee^{-\beta\Hop^I} \right]. \label{partitionsplit}
\end{equation}
Using the definition of a canonical density operator
$\dor_\text{can}\equiv
\tfrac{\ee^{-\beta\Hop}}{Z}$ we rewrite
\begin{equation}
Z(t) = \text{Tr}\!\left[\dor^S_\text{can}(t)\otimes\dor^C_\text{can}
  \ee^{-\beta\Hop^I} \right] Z^S(t)Z^C. \label{partitionsplit_density}
\end{equation}
For the time-dependency of the trace we find based on the Liouville
von-Neumann equation
\begin{equation}
\frac{\dd}{\dd t} \left[\frac{Z(t)}{Z^S(t)Z^C}\right] = -\iu
\text{Tr}\biggl\{\Kom{\Hop(t')}{\dor^S_\text{can}(t)\otimes\dor^C_\text{can}}
  \ee^{-\beta\Hop^I}\biggr\}.
\end{equation}
Making use of the cyclic property of the trace
and the commutator relations~(\ref{kom_microcanonical}), we end up with
\begin{equation}
\frac{\dd}{\dd t} \text{Tr}\!\left[\dor^S_\text{can}(t)\otimes\dor^C_\text{can}
  \ee^{-\beta\Hop^I}\right] = 0 \; \forall \; t. \label{traceconstant}
\end{equation}
Thus, we particularly have
\begin{equation}
\text{Tr}\!\left[\dor^S_\text{can}(t)\otimes\dor^C_\text{can}
  \ee^{-\beta\Hop^I}\right] =
\text{Tr}\!\left[\dor^S_\text{can}(0)\otimes\dor^C_\text{can}
  \ee^{-\beta\Hop^I}\right]. 
\end{equation}
From this we conclude using Eq.~(\ref{partitionsplit_density}) in
Eq.~(\ref{eq:jarzynski_2}) 
\begin{equation}
  \overline{\ee^{-\beta W}} = \overline{\ee^{-\beta W^S}} = \frac{Z^S(t)}{Z^S(0)}. \label{eq:jarzynski_micro}
\end{equation}
The work on the compound system equals that on the open one since no
energy exchange between the system and the environment is allowed and
therefore the substitution in Eq.~(\ref{eq:jarzynski_micro}) on the
left-hand side is justified. Thus the Jarzynski equation holds for
microcanonical coupling.

\subsection{Canonical Coupling - Constant Interaction Energy}
Next, we turn to systems which are allowed to exchange energy with
their surroundings. The Hamilton operator is again of the
form~(\ref{def_hamilton}). We assume the interaction energy to be
a constant of motion:
\begin{equation}
\Kom{\Hop(t)}{\Hop^I} = 0 \; \Rightarrow \;
\Kom{\Hop^S(t)+\Hop^C}{\Hop^I} = 0. \label{kom_canonical}
\end{equation}
This assumption is motivated by the fact that, for a sufficiently
clear separation between system and environment, the interaction
energy should be very small compared to the energy of the system and
the environment, respectively. Therefore, its change should also be
considered to be negligible.  The proof then runs in complete analogue
to that of the microcanonical coupling scenario. We start using the
Jarzynski relation~(\ref{jarzynski}) for closed systems. The partition
sum can be written as in Eq.~(\ref{partitionsplit_density}) using the
commutator relation~(\ref{kom_canonical}):
\begin{equation}
Z(t) = \text{Tr}\!\left[\ee^{-\beta(\Aop+\Bop)} \right]
\overset{(\ref{kom_canonical})}{=}
  \text{Tr}\!\left[ \ee^{-\beta\Aop}\ee^{-\beta\Bop} \right], 
\end{equation}
with $\Aop\equiv \Hop^S(t)\otimes\op{1}^C + \op{1}^S\otimes\Hop^C$ and
$\Bop\equiv \Hop^I$. Using $\kom{\Aop}{\Bop}=0$ and re-substituting
$\Aop$ and $\Bop$ we easily arrive at 
\begin{equation}
Z(t) = \text{Tr}\!\left[\dor^S_\text{can}(t)\otimes\dor^C_\text{can}
  \ee^{-\beta\Hop^I} \right] Z^S(t)Z^C \label{eq:partitionsum_decomposition}
\end{equation}
as in the case of microcanonical coupling. Again, we investigate the
time dependency of the trace using the Liouville von-Neumann equation
and the cyclic property of the trace,
\begin{equation}
\frac{\dd}{\dd t}\left[\frac{Z(t)}{Z^S(t)Z^C}\right] = -\iu
\text{Tr}\biggl\{\dor^S_\text{can}(t)\otimes\dor^C_\text{can}
  \Kom{\ee^{-\beta\Hop^I}}{\Hop(t')}\biggr\}.
\end{equation}
Using the commutator relation~(\ref{kom_canonical}) we conclude that
$\kom{\ee^{-\beta\Hop^I}}{\Hop(t')}= 0 \;\forall\; t'$. Thus, the
relation~(\ref{traceconstant}) also holds for this scenario and
therefore we conclude using Eq.~(\ref{eq:partitionsum_decomposition})
and relation~(\ref{traceconstant}) in Eq.~(\ref{jarzynski}) that we still have
\begin{equation}
\overline{\ee^{-\beta W}} = \frac{Z^S(t)}{Z^S(0)}. \label{eq:jarzynski_const_int}
\end{equation}
This proves that the Jarzynski relation also holds for canonical
coupling which conserves the interaction energy.

Note that the work $\Delta W$ performed on the compound closed system
equals the work $\Delta W_\text{sub}$ done on the subsystem S, \ie\
$\Delta W=\Delta
W_\text{sub}$~\cite{pre_quantum_masterequation_jarzynski}. This is, in
general, different from the energy change of the subsystem during this
process since the subsystem also exchanges energy with its environment
\begin{equation}
\Delta E_\text{sub} = \Delta W_\text{sub} + \Delta Q_\text{sub},
\end{equation}
whereas the energy change of the total, closed system still equals the
work performed on it:
\begin{equation}
\Delta E_\text{tot} = \Delta W
\end{equation}
So, the relation~(\ref{eq:jarzynski_const_int}) is applicable if one
would, in an experiment, measure $\Delta W=\Delta W_\text{sub}$, which
is the work one has to supply in order to perform a chosen process.

So, Eq.~(\ref{eq:jarzynski_const_int}) can be rewritten as
\begin{equation}
\overline{\ee^{-\beta W^S}} =  \frac{Z^S(t)}{Z^S(0)}.
\end{equation}

\subsection{Canonical Coupling - High temperature limit \label{highT}}
Finally, we consider systems with a relatively high temperature $T$, \ie\
we have a small inverse temperature $\beta$ at the beginning. We can
then expand the exponent of the partition sum according to the
Baker-Campbell-Hausdorff formula and related identities. In the
following, we will use the following abbreviations:
\begin{equation}
A \equiv \Hop^S(t)\otimes\op{1}^C + \op{1}^S\otimes\Hop^C \text{ and }
B \equiv \sigma \Hop^\text{int}.
\end{equation}
We thus have
\begin{equation}
\ee^{-\beta A}\ee^{-\beta B} = \ee^{-\beta (A+B) +
  \frac{\beta^2}{2}\kom{A}{B}} + \ord{\beta^3},
\end{equation}
which can be rewritten as
\begin{equation}
\ee^{-\beta A}\ee^{-\beta B} = \ee^{-\beta (A+B)}
\ee^{\frac{\beta^2}{2}\kom{A}{B}}. \label{Eab}
\end{equation}
On the other hand,
\begin{equation}
\ee^{-\beta B}\ee^{-\beta A} =  \ee^{-\beta (B+A)}
  \ee^{\frac{\beta^2}{2}\kom{B}{A}} + \ord{\beta^3}. \label{Eba}
\end{equation}
We combine Eq.~(\ref{Eab}) with Eq.~(\ref{Eba}) and using
\begin{equation}
\Kom{\ee^{-\beta (B+A)}}{\ee^{\frac{\beta^2}{2}\kom{B}{A}}} = 0
+ \ord{\beta^3}
\end{equation}
we arrive at
\begin{align}
\underset{(\ref{Eab})}{\underbrace{\ee^{-\beta A}
    \ee^{-\beta B}}} \underset{(\ref{Eba})}{\underbrace{\ee^{-\beta B}
    \ee^{-\beta A}}} & =  \ee^{-\beta (A+B)}
\ee^{\frac{\beta^2}{2}\kom{A}{B}}  \ee^{\frac{\beta^2}{2}\kom{B}{A}}
\ee^{-\beta (B+A)} \nonumber \\ 
& = \ee^{-2\beta(A+B)}. \label{Etrenn2beta}
\end{align}
Thus, we find:
\begin{align}
  Z(t) = \text{Tr}\!\left[\ee^{-\beta(A+B)}\right] &
  \overset{(\ref{Etrenn2beta})}{=}
  \text{Tr}\!\left[\ee^{-\frac{\beta}{2} A}\ee^{-\frac{\beta}{2}
      B}\ee^{-\frac{\beta}{2} B}\ee^{-\frac{\beta}{2} A}\right]
  \nonumber \\
  & \overset{\hphantom{(\ref{Etrenn2beta})}}{=}
  \text{Tr}\!\left[\ee^{-\beta A} \ee^{-\beta B}\right],
\end{align}
or, re-substituting $A$ and $B$:
\begin{equation}
Z(t) = \text{Tr}\!\left[\ee^{-\beta\left(\Hop^S(t)\otimes\op{1}^C +
  \op{1}^S\otimes\Hop^C\right)}\ee^{-\beta\Hop^I}\right].
\end{equation}
We now can proceed as above:
\begin{equation}
Z(t) = \text{Tr}\!\left[
  \op{\varrho}^S(t)\otimes\op{\varrho}^C\ee^{-\beta\Hop^I}\right] Z^S(t)Z^C. 
\end{equation}
We again investigate the time-dependency of the trace using the Liouville
von-Neuman equation:
\begin{align}
  \frac{\dd}{\dd t} & \left[ \frac{Z(t)}{Z^S(t)Z^C} \right] = -\iu
  \text{Tr}\!\left\{
    \komm{\Hop(t')}{\op{\varrho}^S(t)\otimes\op{\varrho}^C}\ee^{-\beta
      \Hop^I}  \right\} \nonumber \\
  & = -\iu \text{Tr}\!\left\{ \komm{\Hop^S(t')\otimes\op{1}^C +
      \op{1}^S\otimes\Hop^C}{\op{\varrho}^S(t)\otimes\op{\varrho}^C}\ee^{-\beta
      \Hop^I}  \right\} \nonumber \\
  & = -\iu \text{Tr}\!\left\{
    \komm{\Hop^S(t')\otimes\op{1}^C}{\op{\varrho}^S(t)\otimes\op{\varrho}^C} \ee^{-\beta  \Hop^I} \right\} \nonumber \\
  & = -\iu \text{Tr}\!\left\{
    \komm{\Hop^S(t')}{\op{\varrho}^S(t)}\otimes\op{\varrho}^C
    \ee^{-\beta \Hop^I} \right\},
\end{align}
where we have used the cyclic property of the trace in the first line
in order to get rid of the interaction term of the Hamiltonian. If we
have $\Kom{\Hop^S(t')}{\Hop^S(t)} = 0 \; \forall \; t'$, as for a
time-dependent Zeeman splitting of a spin, \eg , we immediately
arrive at
\begin{equation}
\frac{\dd}{\dd t} \left[ \frac{Z(t)}{Z^S(t)Z^C} \right] = 0
\end{equation}
and therefore
\begin{equation}
\frac{Z(t)}{Z(0)} = \frac{Z^S(t)}{Z^S(0)} = \overline{\ee^{-\beta W}}
= \overline{\ee^{-\beta W^S}}. \label{eq:jarzynski_open_high_beta}
\end{equation}
So, for high initial temperatures of the open quantum system, the Jarzynski
relation holds if the Hamiltonian of the system commutes at different
times. This can be shown to hold also in third order of $\beta$, see
appendix~\ref{highT3rd}. 

From the derivation of this result one might expect that the Jarzynski
relation would not hold in this form for arbitrary $\beta$, but that it
could fail and that the deviation of the average
$\overline{\ee^{-\beta W}}$ from the relation of the partition sums
should become larger the smaller the inverse temperature $\beta$. This
expectation will be tested numerically. 
\section{Numerical Results \label{numeric}}

\begin{figure}
  \begin{center} \scalebox{0.7}{
     \includegraphics[]{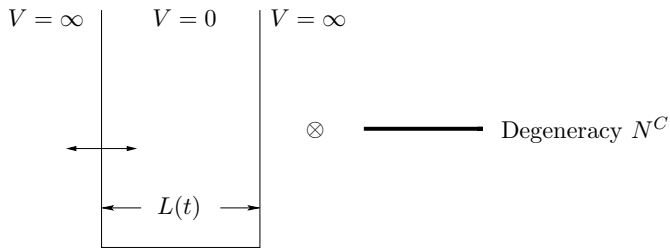}
     }
     \caption{Model system for our numerical investigations. On
       the left-hand side the quantum well with a movable wall is
       depicted. The well is coupled to an environment with a single
       energy level of high degeneracy. \label{fig_model_system}}
     \end{center}
\end{figure}
As a simple model system we choose the standard potential well with
width $L$ containing one particle (particle in a box). Its spectrum is
truncated at level $n$ (cut-off). The well is coupled
\emph{microcanonically} to an environment (cf.
Fig.~\ref{fig_model_system}). To ensure that no energy is exchanged
between system and environment, we model the surroundings as a single,
highly degenerate, energy
level~\cite{book_gemmer_quantum_thermodynamics}.  

One wall of this well (the left one, say) is now supposed to be
movable, allowing for dilation and compression of the well [well width
$L(t)$]. This time-dependence will model the working process. The
particle is prepared initially in a canonical state. Then the
time-evolution of the system is calculated numerically under pure
Schr\"{o}dinger dynamics. From this we can infer the work
distribution for any control function $L(t)$. The discrete values of
$W$ result from the discrete spectral energy differences. The
non-monotonous behavior of $P(W)$ is a combined effect of the spectrum
and the initial occupation probabilities. We use this $P(W)$ in order
to compute the average $\overline{\ee^{-\beta W}}$. This result is
then compared with the relation of the partition sums
$\tfrac{Z^S(t)}{Z^S(0)}$, which can easily be computed independently
for this model. We define $\overline{\ee^{-\beta W}} =
\gamma_\text{J}\tfrac{Z^S(t)}{Z^S(0)}$ with $\gamma_\text{J}$ being a
test factor. If the Jarzynski relation is correct, the test factor
should always equal $1$ within numerical accuracy.

Since we want to consider the work done on the compound, closed
system, we define the work as $W \equiv \Delta E$, since the work
simply equals the energy change of the closed system. Since we will,
in general, find the system in different states for each measurement,
the work is said to fluctuate. However, the definition of work and the
corresponding fluctuations is still a matter under dispute
\cite{epl:work_definition_engel}.

\subsection{Closed Systems}
For our investigation of the closed system we choose the interaction
strength with the environment to be zero.

\begin{figure}
  \begin{center} \scalebox{0.7}{    
    \includegraphics{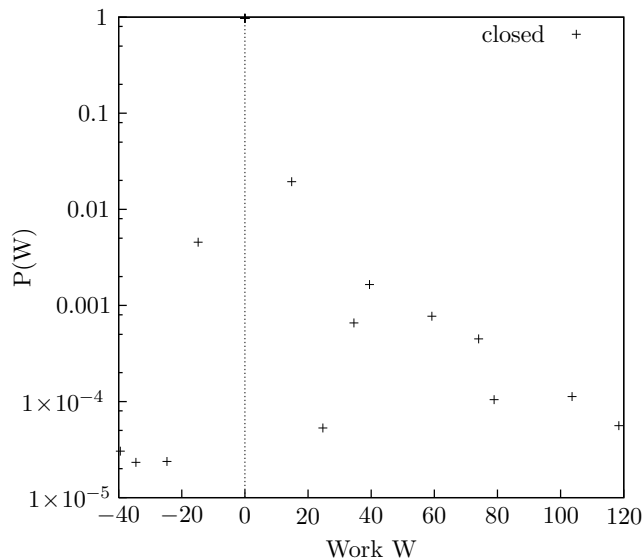}
    }

    \caption{Distribution of work P(W) for a closed system. A circular
      process with $\Hop(t_\text{final})=\Hop(0)$ and a maximal width of
      $L_\text{max}=2L(0)$ was chosen. The wall has been moved with
      constant velocity $\pm v$. The velocity of the process as
      well as the unit energy is in arbitrary units. Probabilities
      smaller than $1\!\times\! 10^{-5}$ are suppressed; cutoff at
      $n=5$ is justified by a low initial temperature. The average work
      $\overline{W}$ is larger than zero. There is a finite chance to
      gain work while performing this process, as can be seen 
      from the points to the left of the dashed line $W=0$.
      \label{fig:numericalresultspotwellclosed_closed}}
  \end{center}
\end{figure}

Our numerics show that, within numerical errors, for any process
realization and any process velocity chosen, we, indeed, have
$\gamma_\text{J}=1$ which means that the Jarzynski relation is found
to be valid. This result holds for different wall displacement
velocities, different initial temperatures or different well widths. See
Fig.~\ref{fig:numericalresultspotwellclosed_closed} for a typical
result for the work distribution obtained.

\subsection{Microcanonical Coupling}
Next, we turn to systems with $\sigma \neq 0$. In order to check
whether the Jarzynski relation holds, we need to consider a process for
which the final Hamiltonian differs from the initial one; otherwise
one trivially would have $\tfrac{Z(t)}{Z(0)} =
\tfrac{Z^\mathrm{S}(t)}{Z^\mathrm{S}(0)} = 1$. For now, we focus on
expansions of the quantum well.

\begin{figure}
  \begin{center} \scalebox{0.7}{    
    \includegraphics{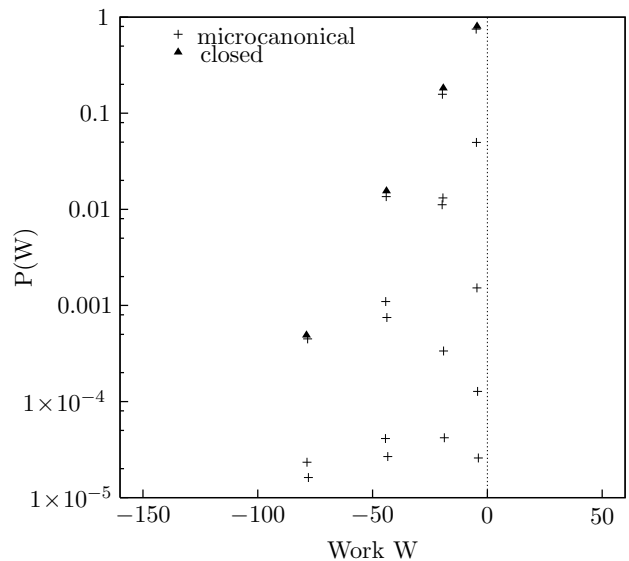}
    }

    \caption{Distribution of work P(W) for a closed and a
      microcanonical system, respectively. An  expansion with 
      $L(t_\mathrm{final})=2L(0)$ was chosen. The wall has been moved with
      constant velocity $v$. The degeneracy was $N^C=150$ and the
      coupling strength $\sigma=0.015$ for the microcanonical
      coupling. For the closed system, $\sigma$ was set
      to zero. The velocity of the process as
      well as the unit energy is in arbitrary units. Probabilities
      smaller than $1\!\times\! 10^{-5}$ are suppressed; cutoff at $n=5$. 
      \label{fig:potwell_micro_closed}}
  \end{center}
\end{figure}

We find that the work distribution is slightly affected by the
microcanonical environment (cf. Fig.~\ref{fig:potwell_micro_closed}).
For better comparison the work distribution for the closed system is
depicted, too. We get $\gamma_\mathrm{J} = 1$ within our numerical
accuracy, indicating that the Jarzynski relation is valid.

This also holds when changing the coupling strength, enlarging the
environment by increasing the degeneracy, or changing the initial
temperature. Note that the distribution of work for the closed system
is, indeed, though only slightly, different from the one obtained for
the same boundary conditions under microcanonical coupling.

Due to the interaction between system and environment the energy
levels of the system are "smeared out". Despite the fact that the
interaction is weak, the system's energy levels are each splitted into
many energy levels with slightly different energy eigenvalues
according to the number of energy levels in the environment and their
respective degeneracies. Some of these energy levels are resolved
within numerical accuracy leading to slightly distinct values of work
$W$. This might lead to the impression that we have multiple values of
work probabilities $P(W)$ for a given work $W$ in
Fig.~\ref{fig:potwell_micro_closed} to Fig.~\ref{fig:potwell_can_hot}
for the open system. This is not the case since for exactly - in the
numerical sense - equal work values, the probabilities add up to a
single probability $P(W)$.

\subsection{Canonical Coupling}
\begin{figure}
  \begin{center} \scalebox{0.7}{    
    \includegraphics{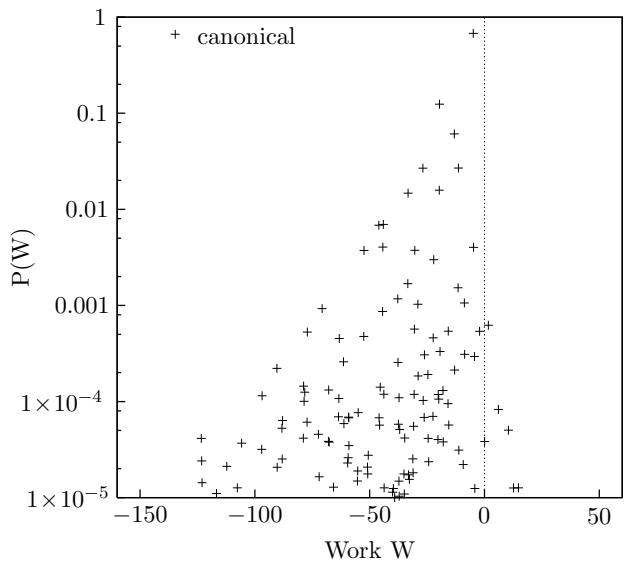}
    }
    \caption{Distribution of work P(W) for a canonical system. An
      expansion with $L(t_\mathrm{final})=2L(0)$ and $\sigma 
      = 0.015$ was chosen. The wall has been moved with
      constant velocity $v$.  The velocity of the process as
      well as the unit energy is in arbitrary units. Probabilities
      smaller than $1\!\times\! 10^{-5}$ are suppressed; cutoff at $n=5$. 
      \label{fig:potwell_can}}
  \end{center}
\end{figure}

We finally turn to a canonical bath, investigating whether the
Jarzynski relation still holds in the high temperature limit, as
predicted. To this end, we couple a canonical bath to the quantum
well. It is modeled in such a way, that it allows every possible
transition in our quantum well when in
resonance~\cite{book_gemmer_quantum_thermodynamics}. Its degeneracy
grows with increasing energy. For now, we focus on expansions of the
quantum well, which is effectively cooled down by these dilations. This
will cause energy to flow from the bath into the system trying to
restore the initial temperature and we expect that transitions to
higher energy levels will be facilitated. First, we have chosen a
moderate initial temperature, such that only the first two energy
levels are significantly occupied, resulting typically in a work
distribution as depicted in Fig.~\ref{fig:potwell_can}.

\begin{figure}
  \begin{center} \scalebox{0.7}{    
    \includegraphics{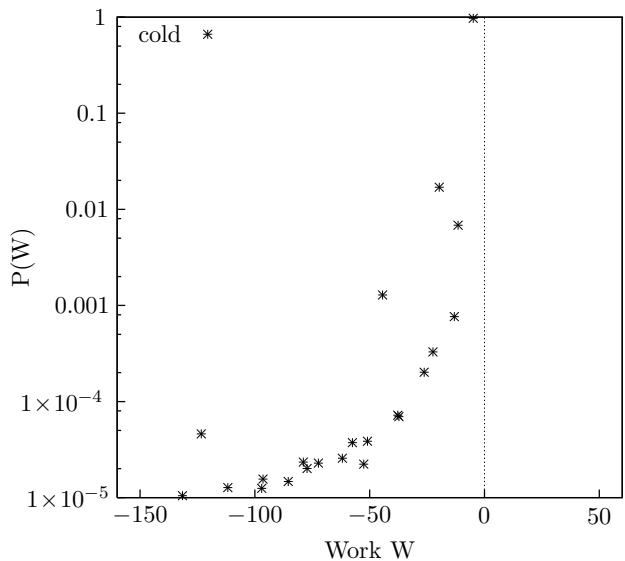}
    }
    \caption{Same as Fig.~\ref{fig:potwell_can}, but with low initial
      temperature \label{fig:potwell_can_cold}}
  \end{center}
\end{figure}

We find that Eq.~(\ref{eq:jarzynski_open_high_beta}) does not hold,
the Jarzynski relation of this form is not strictly valid. We have for
our test factor $\gamma_\mathrm{J} = 0.999$, which quantifies the
still small deviation from the Jarzynski relation. From
sect.~\ref{highT} we suppose that the lower the temperature the bigger
this deviation.  Indeed, if we choose the initial temperature to be
very low, such that almost only the ground state is occupied, we have
$\gamma_\mathrm{J} = 0.98$. Moreover, we expect the work distribution
to be very narrow since transitions from the ground state into higher
ones are relatively unlikely (cf. Fig.~\ref{fig:potwell_can_cold}).

\begin{figure}
  \begin{center} \scalebox{0.7}{    
    \includegraphics{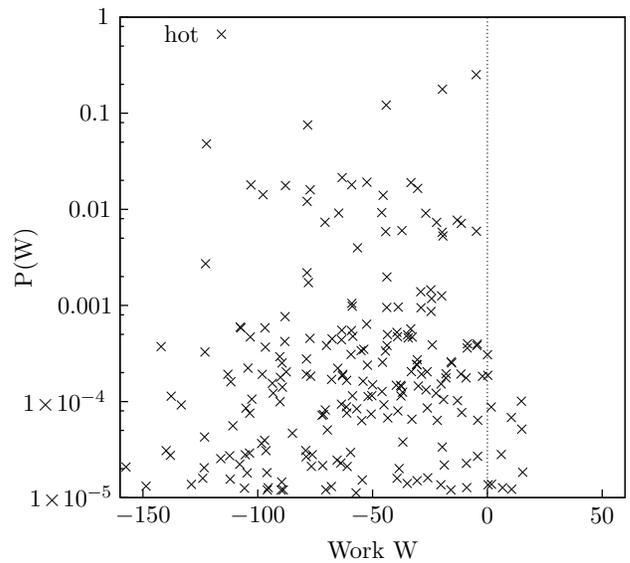}
    }
    \caption{Same as Fig~\ref{fig:potwell_can}, but with high initial
      temperature  \label{fig:potwell_can_hot}}
  \end{center}
\end{figure}

Next, we choose a very high initial temperature, such that almost
every energy level has the same probability to be occupied, in order
to check whether the Jarzynski relation is valid as predicted. The
work distribution is, as expected, rather broad for these
temperatures, cf. Fig.~\ref{fig:potwell_can_hot}. Nevertheless, we
find that $\gamma_\mathrm{J} = 1$ within numerical accuracy. This
means that Eq.~(\ref{eq:jarzynski_open_high_beta}) is fulfilled for
high initial temperatures, confirming our high-temperature limit.

\section{Conclusion}
Starting from the quantum version of the Jarzynski relation for closed
systems, we have shown that this relation can be generalized to open
quantum systems of different coupling types. The key idea has been to split
the exponent of the partition sum into parts for the system and
environment, respectively. This is, in general, not possible, since
the system and environmental Hamiltonian do not commute with the
interaction part. This fact would result in additional terms containing the
commutator of these Hamiltonians.

Nevertheless, we have been able to split up the partition function
under the selected conditions, since then the correction terms cancel.

We have stressed that the problem of fluctuation is intimately related
to some measurement scheme. Further investigations should be guided by
the observation that in quantum mechanics it seems more natural to
consider the fluctuation of a (subsystem-)energy rather than of work
proper~\cite{pre:work_not_observable}.

\appendix
\section{Third order expansion in terms of $\beta\ll 1$  \label{highT3rd}}
We want to extend the result of sect.~\ref{highT} to the third power
of $\beta$. We proceed analogously as above, trying to split up the
exponential function. For $\ord{\beta^4} = 0$ we arrive at
\begin{equation}
\ee^{-\beta A}\ee^{-\beta B} = \ee^{-\beta (A + B) + \frac{\beta^2}{2}\kom{A}{B}}
\ee^{-\frac{\beta^3}{12}\bigl(\dkommAneu + \dkommBneu \bigr)} \label{Eab3}
\end{equation}
and
\begin{equation}
\ee^{-\beta B}\ee^{-\beta A} =
\ee^{\frac{\beta^3}{12}\bigl(\dkommAneu + \dkommBneu \bigr)}
\ee^{-\beta (B + A) - \frac{\beta^2}{2}\kom{A}{B}}. \label{Eba3}
\end{equation}
Combining Eq.~(\ref{Eab3}) with Eq.~(\ref{Eba3}) we have
\begin{align}
\underset{(\ref{Eab3})}{\underbrace{\ee^{-\beta A} \ee^{-\beta B}}}
\underset{(\ref{Eba3})}{\underbrace{\ee^{-\beta B} \ee^{-\beta A}}} & =
\ee^{-\beta (A + B) + \frac{\beta^2}{2}\kom{A}{B}} \ee^{-\beta (B + A)
  - \frac{\beta^2}{2}\kom{A}{B}} \nonumber \\
& \equiv \ee^{-\beta C_+}\ee^{-\beta C_-}, \label{Eabba}
\end{align}
with  $C_+ \equiv -\beta (A + B) + \frac{\beta^2}{2}\kom{A}{B}$ and
$C_- \equiv -\beta (A + B) - \frac{\beta^2}{2}\kom{A}{B}$. We
investigate the resulting product introducing the abbreviation
$\kom{C_+}{C_-} \equiv D_\pm$:
\begin{align}
& \ee^{-\beta C_+}\ee^{-\beta C_-} = \nonumber \\
& \ee^{-\beta(C_+ + C_-)
  +\frac{\beta^2}{2} D_\pm - \frac{\beta^3}{12}\bigl( \Kom{C_+}{D_\pm} +
  \Kom{C_-}{D_\pm}\bigr)} + \ord{\beta^4}.
\end{align}
Since $C_+ = C_- + \beta^2\kom{A}{B}$ we have 
\begin{align}
& D_\pm = -\beta^3 \Bigl\{ \dkommAneu + \dkommBneu \Bigr\} \\
& \Rightarrow \kom{C_\pm}{D_\pm} \overset{\ord{\beta^4}}{=} 0 
\end{align}
This leaves us with
\begin{align}
\ee^{-\beta C_+}\ee^{-\beta C_-} & = \ee^{-\beta (C_+ + C_-)
  +\frac{\beta^2}{2} D_\pm} \nonumber \\
& = \ee^{-2\beta (A+B)}\ee^{\frac{\beta^2}{2}D_\pm} + \ord{\beta^4} 
\end{align}
and analogously
\begin{equation}
\ee^{-\beta C_-}\ee^{-\beta C_+} =
\ee^{\frac{\beta^2}{2}D_\mp}\ee^{-2\beta (A+B)} + \ord{\beta^4}, 
\end{equation}
with $D_\mp \equiv \kom{C_-}{C_+} = - D_\pm$. Combining these two
relations we have
\begin{equation}
\ee^{-\beta C_+}\ee^{-\beta C_-} \ee^{-\beta C_-}\ee^{-\beta C_+} =
\ee^{-4\beta (A+B)}. \label{Ec+c-c-c+}
\end{equation}
Therefore we have
\begin{align}
\text{Tr}\! \left( \ee^{-\beta A}\ee^{-\beta B} \right) &
\overset{\hphantom{(\ref{Eabba})}}{=} \text{Tr}\! \left(
  \ee^{-\frac{\beta}{2} A}\ee^{-\frac{\beta}{2} B}
  \ee^{-\frac{\beta}{2} B} \ee^{-\frac{\beta}{2} A}  \right) \nonumber \\
& \overset{(\ref{Eabba})}{=} \text{Tr}\!\left( \ee^{\frac{-\beta}{2}C_+}
\ee^{\frac{-\beta}{2}C_-}\right) \nonumber \\
& \overset{\hphantom{(\ref{Eabba})}}{=} \text{Tr}\!\left(
  \ee^{\frac{-\beta}{4}C_+} \ee^{\frac{-\beta}{4}C_-}\ee^{\frac{-\beta}{4}C_-}
\ee^{\frac{-\beta}{4}C_+}   \right) \nonumber \\
& \overset{(\ref{Ec+c-c-c+})}{=} \text{Tr}\!\left( \ee^{-\beta(A+B)}
\right) = Z.
\end{align}
The rest of the proof for the Jarzynski relation to hold runs
analogously to sect.~\ref{highT}.
%
%
%
%
%

%
%
%
\end{document}